\documentclass{aa}  

\usepackage{graphicx}
\usepackage{txfonts}
\begin{document}

   \title{How likely is the interstellar origin of CNEOS14?}

   \subtitle{On the reliability of the CNEOS database}

 \author{H. Socas-Navarro
          \inst{1,2}
          }

   \institute{Instituto de Astrof\'\i sica de Canarias, V\'\i a L\'actea S/N, La Laguna 38205, Tenerife, Spain\\
   \email{hsocas@iac.es}
         \and
             Departamento de Astrof\'\i sica, Universidad de La Laguna, 38205, Tenerife, Spain \\
             }
             

 
  \abstract{
 This paper investigates the likelihood that the CNEOS 2014-01-08 superbolide (CNEOS14) was caused by an interstellar object. This issue has remained controversial due to lack of information on the capabilities of the classified satellite sensors that recorded the fireball. We critically evaluate previous studies, specifically addressing the reliability of the CNEOS database and the associated measurement uncertainties. With proper statistical analysis of existing data and the addition of a relevant new event (the 2024 Iberian superbolide), we disprove some claims in previous work, such as: a) the existence of a purported correlation between CNEOS velocity errors and bolide speed; b) the presence of large velocity errors of 10-15 km/s in the CNEOS database; and c) the assertion that CNEOS14 is most likely a solar system object with a hyperbolic trajectory due to measurement errors. We present a quantitative estimate of the probability that CNEOS14 is interstellar. If its measurement errors are drawn from the same underlying distribution as the 18 calibrated events, then the probability that CNEOS14 is interstellar is 94.1\%. This probability is lower than the 99.7\% confidence (3-$\sigma$) generally required to claim a scientific discovery. However, it is sufficiently high to be considered significant and, by far, the most likely explanation for the currently available empirical evidence. }



   \maketitle
%

\section{Introduction}

The possibility that some meteors might be of interstellar origin has motivated extensive investigation and searches for such events since at least the 1950s (\citealt{fisher1928hyperbolic,opik1950interstellar,almond1950interstellar,pena2022orbital} and references therein). However, the unambiguous identification of interstellar impactors on our planet has proven to be a formidable challenge (\citealt{hajdukova2020challenge}).

Empirical evidence exists for interstellar objects passing through our solar system. Two sub-kilometer bodies have been observed in unbound trajectories, 1I/'Oumuamua and 2I/Borisov. On the smaller end of the size spectrum, in situ measurements of dust by the spacecraft Ulysses and Galileo have resulted in the unequivocal detection of particles coming from beyond the boundaries of our solar system. Using these two extremes of the size distribution as constraints, one can estimate that the distribution of flux versus particle size may be approximated with a $r^{-3}$ power law (\citealt{jewitt2023interstellar}), where $r$ is the radius of the object. There should be an abundance of interstellar meteoroids of all sizes, even asteroids, routinely impacting our planet. However, except for one notable exception, observational uncertainties have prevented a sufficiently confident separation of hyperbolic and elliptical orbits, thus precluding the identification of interstellar meteors or bolides. 

\cite{siraj2022meteor} identified a peculiar event in the CNEOS fireball database maintained by NASA's Jet Propulsion Laboratory\footnote{https://cneos.jpl.nasa.gov/fireballs/}. The bolide was recorded with a geocentric speed of 44~km~s$^{-1}$, implying that the meteoroid had a very hyperbolic heliocentric speed of 60~km~s$^{-1}$ (the Sun's escape velocity at 1~AU is 42~km~s$^{-1}$). Although the authors referred to it as IM1 (interstellar meteor 1), we shall adopt here the more neutral nomenclature CNEOS14 to remain agnostic about its possible interstellar origin. By that name we shall designate both the fireball and the meteoroid, as the context will make it clear which one we are alluding to.

CNEOS14 exhibits some peculiarities; most notably, it has a low inclination with respect to the plane of the ecliptic (7$^{\rm o}$) and a high velocity with respect to the local standard of rest if we assume that it has not interacted with any other object on its way before reaching our planet. \cite{socas2023candidate} pointed out that these anomalies would be resolved if the meteoroid had experienced a gravitational encounter with the putative Planet Nine, a scenario that is further supported by the coincidence of its trajectory with the highest probability region of finding Planet Nine (\citealt{brown2021orbit}).

Following the claim of an interstellar bolide discovery, the authors organized an expedition to search for a meteorite on the ocean floor near the impact site. The use of infrasound and seismic stations to obtain a precise location (\citealt{siraj2023localizing}), the recovery of spherules, allegedly from CNEOS14, and their analysis (\citealt{loeb2024recovery,loeb2024recoveryb}) were subjects of intense controversy in the scientific literature (\citealt{fernando2024probably,gallardo2023anthropogenic,gallardo2024anthropogenic,desch2024u}), but also in mainstream media, sparked in great measure by the suggestion of an artificial extraterrestrial origin (\citealt{loeb2024recovery}). This controversy extends to a wider context of high-profile claims by these authors on key areas of solar system research, including a cometary origin for the Chicxulub impactor (\citealt{siraj2021breakup}), or the extraterrestrial technological origin of 1I/'Oumuamua (\citealt{bialy2018could,sheerin2021oumuamua,loeb2022possibility}). Other authors deem all these extraordinary claims as unfounded, exaggerated or not backed by empirical data (\citealt{desch2022breakup,bannister2019natural}).

A recent paper by \cite{bb23} (hereafter BB23) raises doubts on the interstellar origin of CNEOS14, stating instead that it was most likely a solar system object. Its apparent interstellar origin, according to these authors, would be the result of a combination of large measurement errors. The CNEOS database is populated with events provided to NASA by the U.S. Space Force which, unfortunately, are obtained by satellite sensors whose characteristics have not been disclosed (\citealt{tagliaferri1994detection}). In particular, the measurement uncertainties are not published. Given the scientific relevance of this particular event, some further information has been disclosed by the U.S. Department of Defense, including the light curve\footnote{https://cneos.jpl.nasa.gov/fireballs/lc/bolide.2014.008.170534.pdf} and a statement confirming (using additional data available to them) that the data in the CNEOS database is sufficiently accurate to indicate the meteoroid's interstellar trajectory\footnote{https://lweb.cfa.harvard.edu/~loeb/DoD.pdf}. Fortunately, some information about the reliability of the measurements may be inferred from a set of CNEOS fireballs that were also measured by independent scientific observatories (see section~\ref{sec:velerr} below).

We quantified the likelihood that, based on the available evidence, CNEOS14 was a solar system object and tested some relevant claims of BB23 that were made without a quantitative justification. Specifically, we find the following claims to be inconsistent with the available evidence: a)there is a correlation between the measured velocity and the velocity error; b)CNEOS14's velocity has an error of 10-15~km~s$^{-1}$; c)the simplest explanation is that CNEOS14 is a solar system object whose trajectory appears hyperbolic due to measurement errors.

Due to their relatively unsurprising nature, these claims are already propagating in the literature as other authors take them as established truth without further scrutiny (\citealt{fernando2024seismic,desch2023critique}). We thus felt compelled to write this paper with the goal of clarifying what can and cannot be inferred from the data in order to prevent further confusion. Debate and even controversy are inherent aspects of the scientific process. However, it is important to maintain it rigorous and dispassionate, and not respond to exaggerated or unfounded claims with others of similar nature. 

\section{Velocity errors}
\label{sec:velerr}

The key factor to assess a possible extrasolar origin of CNEOS14 is its velocity vector. With its measured radiant, the velocity error would need to be greater than 20~km~s$^{-1}$ to make its orbit bound to the Sun (there is, of course, the possibility that a smaller velocity error could be combined with radiant errors; we shall address this in the next section). Even though there is no public information on the CNEOS database uncertainties, it is possible to obtain some upper bounds by comparing data for events that have been simultaneously observed with scientific ground-based facilities. Such comparisons have been performed by \cite{devillepoix2019observation}, \cite{pena2022orbital}, BB23, \cite{pena2024oort}. Similarly to these authors, we shall here adopt the interpretation that the discrepancies with the ground-based measurements are ascribed entirely to errors in the CNEOS entries.

\subsection{Using 2023 data}

\begin{figure*}[ht]
  \centering
    \includegraphics[width=0.45\textwidth]{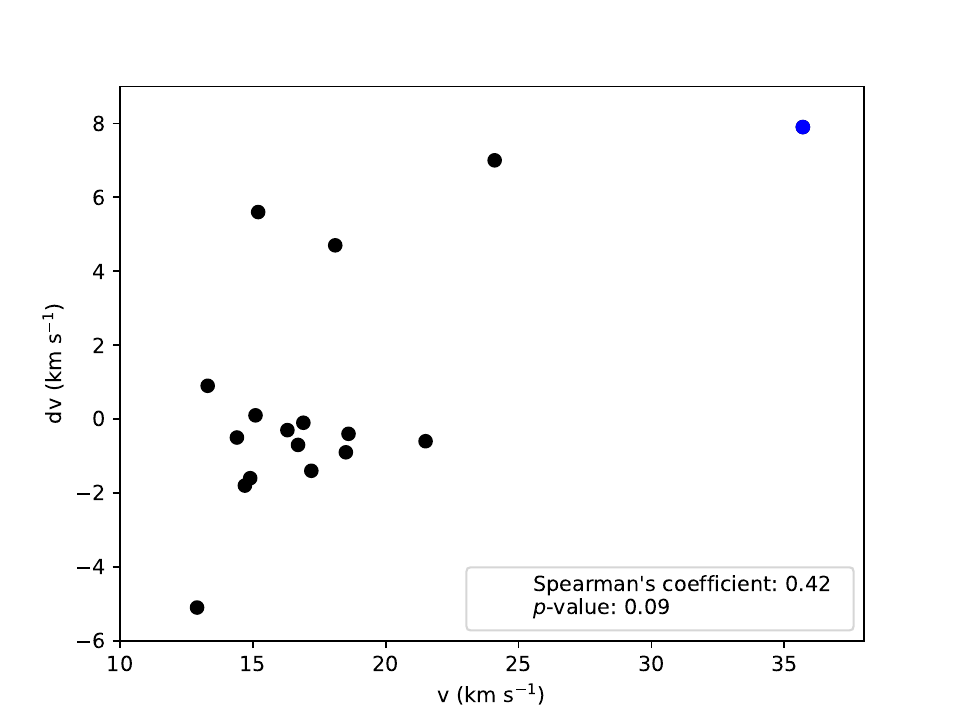}
    \includegraphics[width=0.45\textwidth]{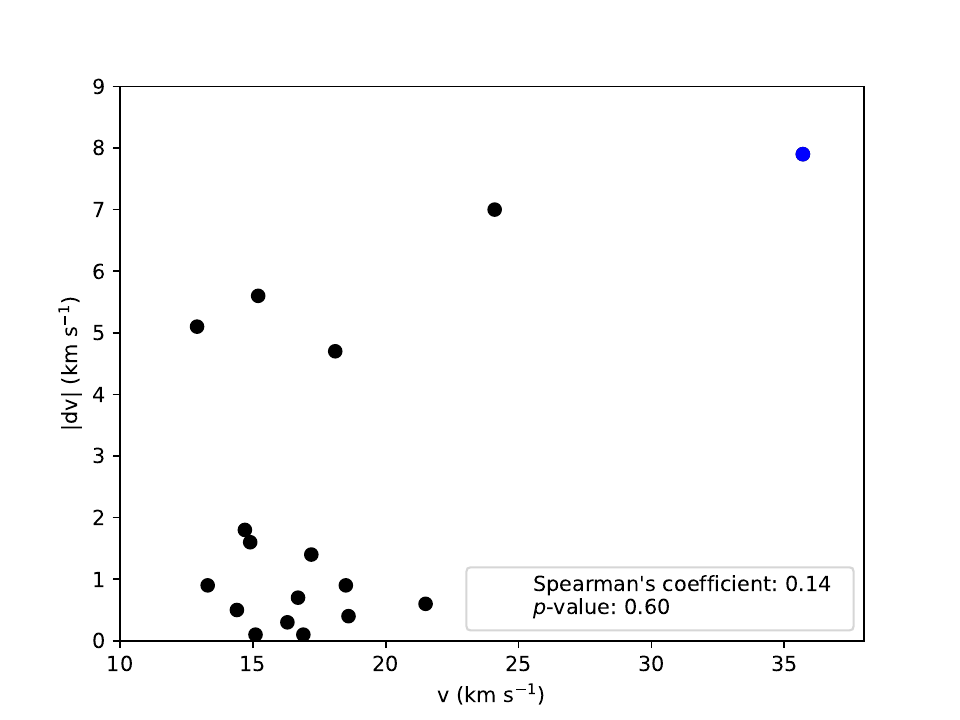}
  \caption{Left panel: Velocity error vs velocity for the 17 events analyzed in BB23. Right panel: Same as the left panel but considering the absolute value of the errors. In both panels, the high-velocity, high-error outlier is marked as a blue point.}
  \label{fig:scatter1}
\end{figure*}

BB23 argued that faster bolides should yield larger measurement errors, which would be true if the measurements were limited by cadence. To test this hypothesis, they plotted the velocity error $dv$ (discrepancy between the CNEOS and ground-based measurements) versus the fireball velocity $v$ for the 17 available events (their Figure 2, recreated here in Figure~\ref{fig:scatter1}, left panel). The result is a rather scattered cloud of points with one outlier (the point marked as a blue dot in Figure~\ref{fig:scatter1}) that has high velocity and high error ($v$=35.7~km~s$^{-1}$, $dv$=7.9~km~s$^{-1}$). The outlier might give the reader a visual impression that there is a correlation between $v$ and $dv$. 

In order to avoid subjectivity and confirmation bias, one needs to quantify the correlation coefficient to determine if a correlation really exists. Unfortunately, BB23 did not make this straightforward calculation and simply relied on the subjective appearance of the plot to make a very strong claim, namely that there is a correlation between the error $dv$ and the velocity $v$. Here we have calculated the relevant numbers to test this assertion. The Spearman's rank correlation coefficient (\citealt{spearman1094proof}) for this particular set of points is a modest $\rho$=0.42 (the presence of a strong outlier makes the Spearman's coefficient a better suited statistical test than the Pearson's coefficient, as it is less sensitive to outliers), with an associated $p$-value of 0.095. In the context of hypothesis contrast, a correlation is considered statistically significant when its associated $p$-value is lower than 0.05. Values greater than 0.10 are generally considered as "no evidence". Therefore, even with the methodology and data employed by BB23, we can conclude that the claim of a correlation between $dv$ and $v$, which they base solely on subjective perception, is not supported by objective evidence.

We observed another problem with the BB23 methodology. In their figure they plotted the velocity errors with sign, thus increasing the influence of the (positive-sign) outlier. In order to rigorously test their hypothesis (that the limited cadence results in larger errors for faster meteors), they should have considered the absolute values of the errors $|dv|$. We show this plot in Figure~\ref{fig:scatter1} (right panel). As expected, taking the absolute value of the errors results in a lower weight of the outlier and even weaker correlation coefficients. The Spearman's coefficient is now 0.14 and its $p$-value is 0.60. In conclusion, there is absolutely no correlation between $|dv|$ and $v$.

\subsection{The 2024 Iberian superbolide}

\begin{figure*}[ht]
  \centering
    \includegraphics[width=0.45\textwidth]{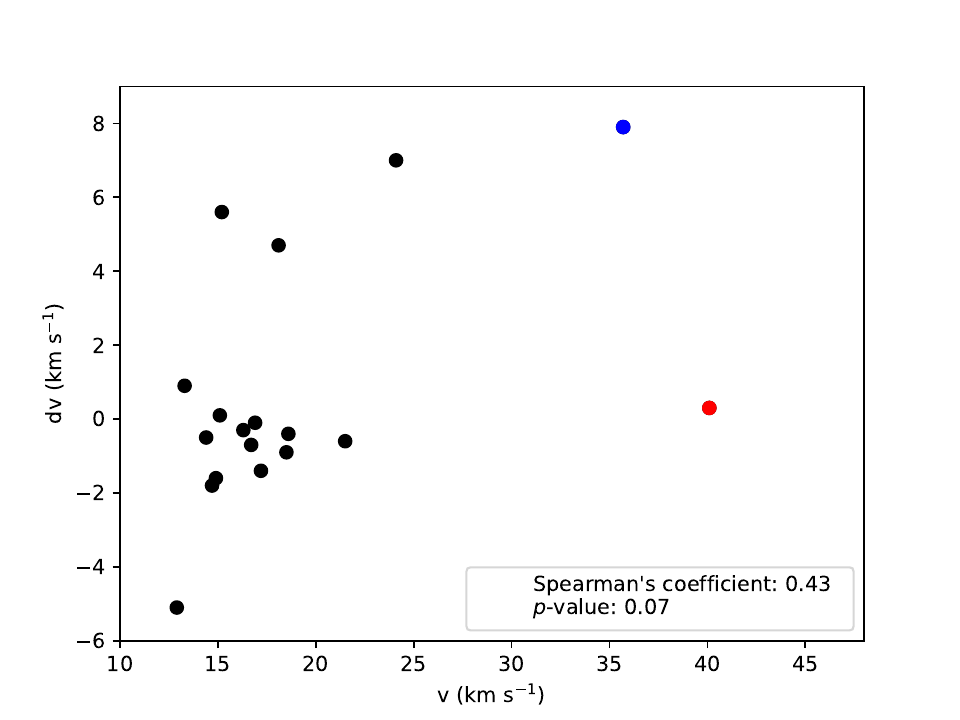}
    \includegraphics[width=0.45\textwidth]{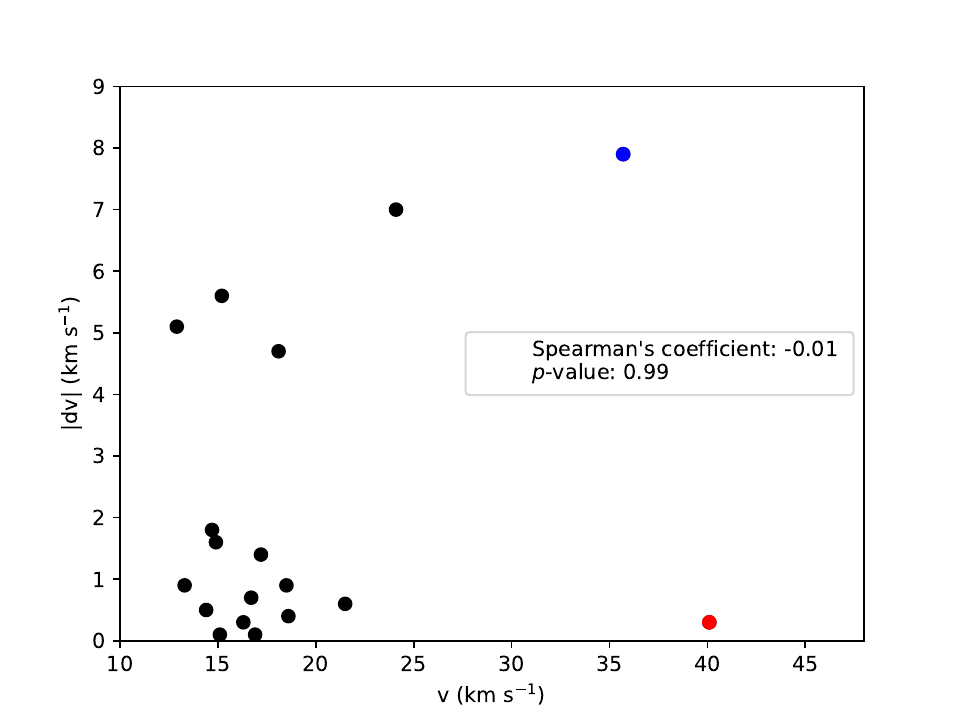}
  \caption{Same as Fig~\ref{fig:scatter1} but including the recent event (2024 Iberian superbolide), shown here as a red point.}
  \label{fig:scatter2}
\end{figure*}

As we have seen, the visual appearance of correlation in the BB23 scatter plot is based entirely on one single outlier point on the far right side of the figure. During the writing of this paper, a very bright fireball occurred that provided very relevant new data. Here we discuss the implications of this fortuitous event. An extremely bright bolide was seen over Spain and Portugal the night of May 18th 2024, which was recorded both by the USG satellite sensors and by scientific observatories (\citealt{epa2024iberian}). The superbolide was nearly as fast as CNEOS14, measured at 40.1~km~s$^{-1}$ by ground-based stations. When it is added to the dataset, we obtain the plots shown in Figure~\ref{fig:scatter2} (the new event is represented by the red dot). As before, the left panel shows the data using the same methodology as BB23, whereas the right panel shows the absolute value of the error, which we believe would be the correct approach. Notice how the addition of this second point on the right-hand side of the figure completely eliminates any possible subjective perception of a correlation between $dv$ and $v$. 

\begin{figure*}[ht]
  \centering
    \includegraphics[width=0.75\textwidth]{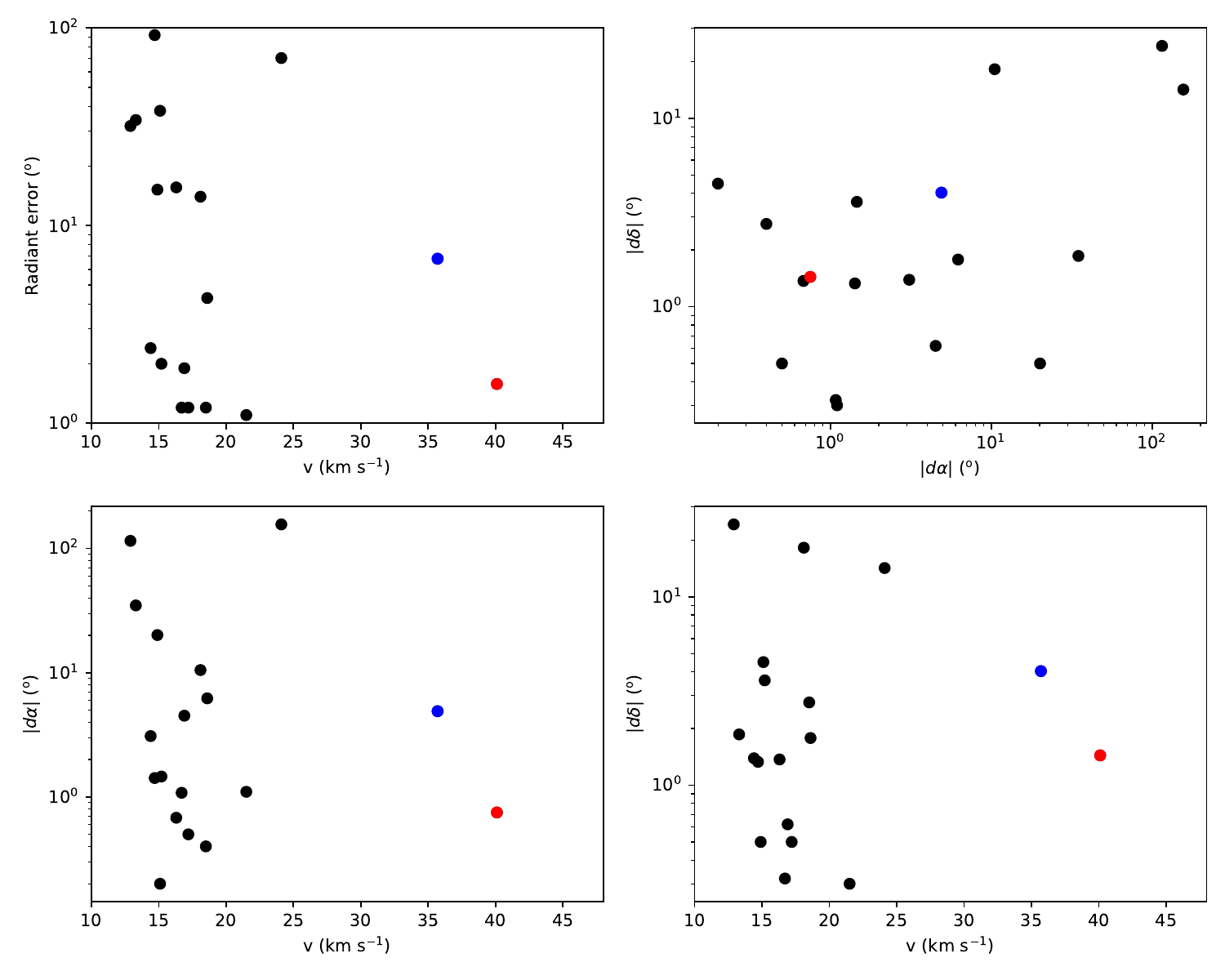}
  \caption{Scatter plots of velocity and radiant errors. The radiant errors are shown combined (upper left panel) or separated into R.A. ($d\alpha$) and dec ($d\delta$). The blue and red points mark the same events as in Figs~\ref{fig:scatter1} and~\ref{fig:scatter2}.}
  \label{fig:radiants}
\end{figure*}

For completeness, Figure~\ref{fig:radiants} shows the radiant errors. The plot in the upper-left panel is similar to Figure~3 in BB23 but updated to include the 2024 Iberian superbolide. The other panels plot the R.A. and declination errors separately.

\subsection{Extrapolation}

There is a second strong claim in BB23 that is not well justified. They state that the comparison of CNEOS data to ground-based observations shows that the former has large uncertainties of 10-15~km~s$^{-1}$ at high-speed events. This is as much as twice the largest value of $dv$ observed in the database. To obtain this 10-15~km~s$^{-1}$ range, they fit a straight line to the points in Figure~\ref{fig:scatter1} and extrapolate it to the CNEOS14 observed speed (44.8~km~s$^{-1}$). However, as we have shown above, there is no significant correlation between $dv$ and $v$ and therefore this fit is meaningless. In fact, if it is done considering the 2024 Iberian superbolide, then the extrapolation yields a completely different (much lower) value.

\begin{figure}[ht]
  \centering
    \includegraphics[width=0.45\textwidth]{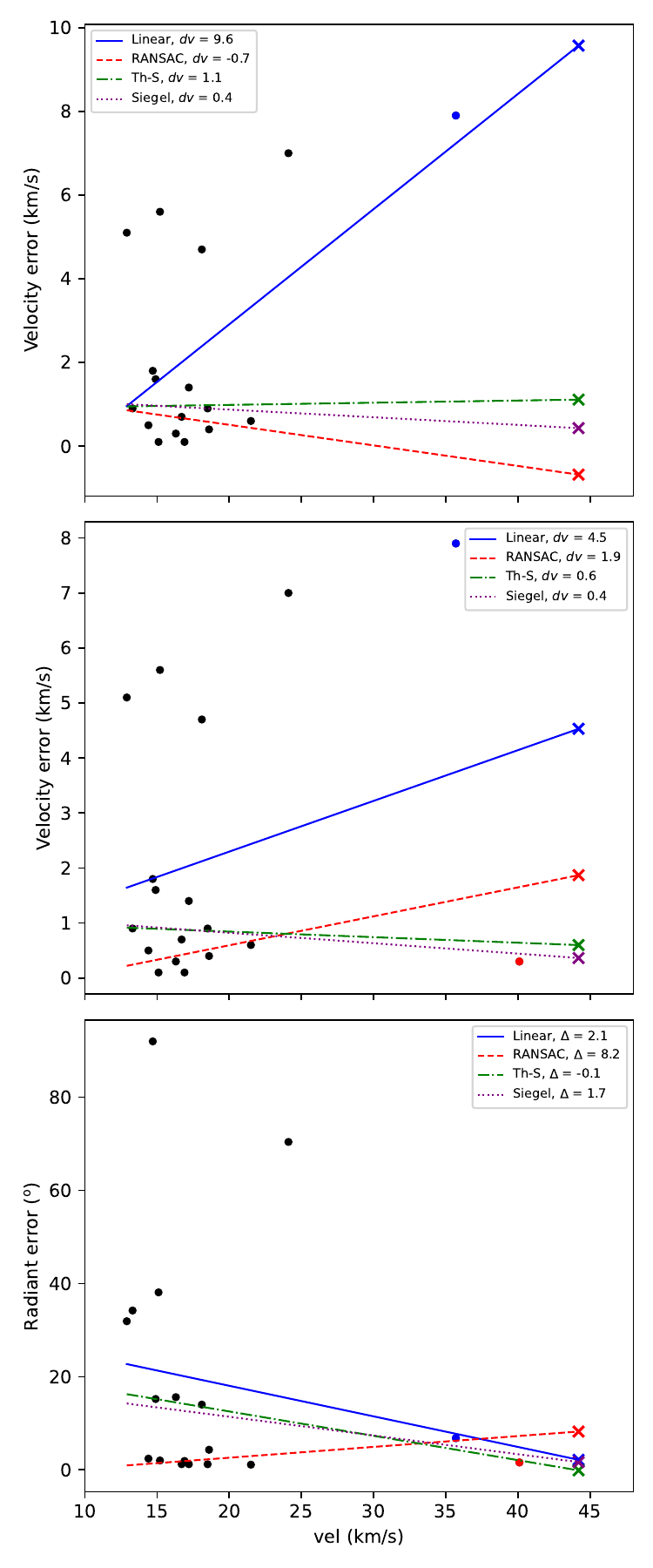}
  \caption{Various fits to various measurement errors as a function of bolide speed and the extrapolation to the error expected for CNEOS14. Top panel: Velocity error for the 17 events analyzed in BB23. Middle panel: Same as the upper panel but including the 2024 Iberian superbolide (red point). Bottom panel: Radiant error. All panels: The blue line is a simple straight line fit, as in BB23. The other lines are three different statistically robust fits, less sensitive to outliers (see text for details and references). The crosses mark the extrapolation of each fit to CNEOS14's speed (numerical values are given in the legend).  }
  \label{fig:fits}
\end{figure}

Figure~\ref{fig:fits} (top panel, blue line) shows the results of fitting the cloud of 17 points to a straight line, as done in BB23 (but with unsigned errors). Clearly, the fit is dominated by a strong outlier (marked by the blue point in the figure), resulting in an extrapolated error $dv$ for CNEOS14 of 9.6~km~s$^{-1}$. We compared this result to what is obtained using three different robust fitting methods: RANSAC (\citealt{ransac}), Theil-Sen (\citealt{theilsen}), and Siegel (\citealt{siegel}). With the robust fits, the extrapolated errors for the CNEOS14 velocity are: -0.7, 1.1 and 0.4~km~s$^{-1}$, respectively. 

If we now add the 2024 Iberian superbolide to the plot (middle panel of Figure~\ref{fig:fits}), we have a second data point (marked in red) in the fast event region of the plot. Even the simple straight line fit now yields a lower extrapolated error of 4.5 km~s$^{-1}$. The robust extrapolations now yield 1.9, 0.6 and 0.4~km~s$^{-1}$, respectively. For completeness, the bottom panel of the figure shows a similar plot for the radiant error.

We conclude that the data of CNEOS errors does not exhibit any correlations that would support the notion of anomalously large errors for CNEOS14, compared to the rest of the dataset. For the 18 events considered here (the 17 in BB23 plus the 2024 Iberian superbolide), the standard deviation of the velocity error is $\sigma$(dv)=3.2~km~s$^{-1}$, the standard deviation of the absolute value of the velocity error is $\sigma$(|dv|)=2.5~km~s$^{-1}$, and the maximum error in the sample is 7.9~km~s$^{-1}$. Figure~\ref{fig:scatter2}
seems to hint at a possible bimodal distribution of errors, with 72\% of events having errors smaller than 2~km~s$^{-1}$ and 28\% between 4~and 8~km~s$^{-1}$, but they are not correlated with speed.

\subsection{Combining velocity and radiant errors}

BB23 argue that, even if the velocity errors are lower than the 20~km~s$^{-1}$ needed to make CNEOS14's orbit unbound, there exist combinations of velocity and radiant errors that would do so. This argument is then used to claim that CNEOS14 is most likely a solar system object. The relevant question in this situation is how likely are such combinations of errors. Evidence for scientific discoveries is generally reported within some confidence intervals (typically 3-$\sigma$ or 99.7\% confidence, but in some scenarios the requirements are even more strict). In order to make the case for dismissing the empirical evidence of CNEOS14's hyperbolic orbit, one would need to show that the probability of the required measurement error is significant. Unfortunately, BB23 did not present any attempts to quantify this probability and relied solely on a qualitative discussion involving a large extrapolated error for CNEOS14 combined with a large radiant error. Figure~\ref{fig:fits} indicates that such scenario might be rather unlikely. In this section we take a quantitative look at the problem.

We have estimated the probability that CNEOS14 is hyperbolic by using the available information on the CNEOS events that have been simultaneously measured by scientific observatories. To that aim, we constructed a matrix with all the measured errors in velocity $dv$ and radiant ($d\alpha$, $d\delta$), where $\alpha$ and $\delta$ represent the R.A. and declination, respectively. Each value of $dv$ and ($d\alpha$, $d\delta$) is included with both positive and negative sign. In total, we have 2$\times$18$\times$2$\times$18 (i.e., 1296) combinations of errors. We then looped through all of these and added them as a perturbation to the nominal speed and radiant of CNEOS14. Each one of these perturbed velocities was then transformed to the solar system barycentric reference frame and we counted how many had a barycentric velocity in excess of the Sun's escape velocity at 1~AU,  42~km~s$^{-1}$. 

We obtained that CNEOS14 is still hyperbolic in 94.3\% of the 1296 cases analyzed. The percentage does not depend significantly on the inclusion of the 2024 Iberian superbolide (93.9\% if we neglect it). This probability is below the 99.7\% that is typically required to claim a firm scientific discovery but it seems sufficient to assert that, based on the best available empirical evidence, CNEOS14 was most likely on a hyperbolic trajectory (assuming that its uncertainties are represented by the same underlying distribution as the 18 calibrated events).

\section{Conclusions}

We have concentrated here on analyzing the velocity vector because, as BB23 noted, it is the crux of the problem. They discussed other arguments, such as the fact that other apparently hyperbolic bolides in the database have orbits that are very close to elliptical, within reasonably small observational errors. This is actually expected behavior (where some highly eccentric orbits may be converted into hyperbolic with small velocity errors) that does not say anything about the probability that CNEOS14 is hyperbolic. In other words, the existence of a population of hyperbolic impostors does not undermine the reality of legitimate hyperbolics. Additionally, arguments concerning the challenges of fitting certain aspects of CNEOS14 to current models are not very persuasive, either. It is plausible that models derived from the fitting of solar system meteors might not apply equally well to a potentially different class of objects with extrasolar origin. Therefore, the real issue is to understand as best as we can the empirical data and the associated uncertainties. In lieu of a disclosure of more detector information, future progress might come from the accumulation of more calibration data in the form of new fireballs simultaneously observed by scientific observatories and the CNEOS sensors. 

In the meantime, the best we can do is to analyze the existing events with adequate statistical methodologies. We have found that, contrarily to assertions made in previous work:
\begin{itemize}
    \item There is no correlation between the CNEOS velocity error and the bolide speed.
    \item There is no reason to claim that the fast CNEOS events have uncertainties of 10-15~km~s$^{-1}$. The maximum error observed in the database is an event with $dv$ = 7.9~km~s$^{-1}$. The standard deviation of the error (in absolute value) is 2.5~km~s$^{-1}$. 
    \item Current empirical evidence indicates that, if the errors of the calibrated set are representative, there is a 94.3\% probability that CNEOS14 is of extrasolar origin.
\end{itemize}

The conclusions of our analysis, initially drawn from previously existing data, are strongly reinforced by the recent 2024 Iberian superbolide. This new event allows us to add a second data point to the fast limit of the calibrated CNEOS population, thus providing a more robust picture of the statistical sample.

\begin{acknowledgements}
The author acknowledges support from the Agencia Estatal de Investigación del Ministerio de Ciencia e Innovación (AEI-MCINN) under grant "Hydrated Minerals and Organic Compounds in Primitive Asteroids" with reference PID2020-120464GB-100. This research has made use of NASA's Astrophysics Data System Bibliographic Services and the NASA JPL CNEOS fireball database.  The Python Matplotlib \citep{H07}, Numpy \citep{numpy11}, Scipy \citep{2020SciPy-NMeth} and IPython \citep{ipython07} modules have been employed to generate the figures and calculations in this paper.

\end{acknowledgements}
%
%

\bibliographystyle{aa}
\bibliography{refs.bib}

\begin{thebibliography}{37}
\expandafter\ifx\csname natexlab\endcsname\relax\def\natexlab#1{#1}\fi

\bibitem[{Akritas {et~al.}(1995)Akritas, Murphy, \& Lavalley}]{theilsen}
Akritas, M.~G., Murphy, S.~A., \& Lavalley, M.~P. 1995, Journal of the American
  Statistical Association, 90, 170

\bibitem[{Almond {et~al.}(1950)Almond, Davies, \&
  Lovell}]{almond1950interstellar}
Almond, M., Davies, J., \& Lovell, A. 1950, The Observatory, Vol. 70, p.
  112-113 (1950), 70, 112

\bibitem[{Bannister {et~al.}(2019)Bannister, Bhandare, Dybczy{\'n}ski,
  Fitzsimmons, Guilbert-Lepoutre, Jedicke, Knight, Meech, McNeill, Pfalzner,
  {et~al.}}]{bannister2019natural}
Bannister, M.~T., Bhandare, A., Dybczy{\'n}ski, P.~A., {et~al.} 2019, Nature
  astronomy, 3, 594

\bibitem[{Bialy \& Loeb(2018)}]{bialy2018could}
Bialy, S. \& Loeb, A. 2018, The Astrophysical Journal Letters, 868, L1

\bibitem[{Brown \& Batygin(2021)}]{brown2021orbit}
Brown, M.~E. \& Batygin, K. 2021, The Astronomical Journal, 162, 219

\bibitem[{Brown \& Borovi{\v{c}}ka(2023)}]{bb23}
Brown, P.~G. \& Borovi{\v{c}}ka, J. 2023, The Astrophysical Journal, 953, 167

\bibitem[{Derpanis(2010)}]{ransac}
Derpanis, K.~G. 2010, Image Rochester NY, 4, 2

\bibitem[{Desch(2024)}]{desch2024u}
Desch, S. 2024, arXiv preprint arXiv:2403.05161

\bibitem[{{Desch} \& {Jackson}(2023)}]{desch2023critique}
{Desch}, S. \& {Jackson}, A. 2023, arXiv e-prints, arXiv:2311.07699

\bibitem[{Desch {et~al.}(2022)Desch, Jackson, Noviello, \&
  Anbar}]{desch2022breakup}
Desch, S.~J., Jackson, A.~P., Noviello, J.~L., \& Anbar, A.~D. 2022, Scientific
  reports, 12, 10415

\bibitem[{Devillepoix {et~al.}(2019)Devillepoix, Bland, Sansom, Towner,
  Cup{\'a}k, Howie, Hartig, Jansen-Sturgeon, \&
  Cox}]{devillepoix2019observation}
Devillepoix, H.~A., Bland, P.~A., Sansom, E.~K., {et~al.} 2019, Monthly Notices
  of the Royal Astronomical Society, 483, 5166

\bibitem[{Fernando {et~al.}(2024)Fernando, Charalambous, Desch, Jackson,
  Mialle, Sansom, \& Ekstr\"om}]{fernando2024probably}
Fernando, B., Charalambous, C., Desch, S., {et~al.} 2024, LPI Contributions,
  3040, 2595

\bibitem[{{Fernando} {et~al.}(2024){Fernando}, {Mialle}, {Ekstr\"om},
  {Charalambous}, {Desch}, {Jackson}, \& {Sansom}}]{fernando2024seismic}
{Fernando}, B., {Mialle}, P., {Ekstr\"om}, G., {et~al.} 2024, arXiv e-prints,
  arXiv:2403.03966

\bibitem[{{Fisher}(1928)}]{fisher1928hyperbolic}
{Fisher}, W.~J. 1928, Harvard College Observatory Circular, 331, 1

\bibitem[{Gallardo(2023)}]{gallardo2023anthropogenic}
Gallardo, P.~A. 2023, Research Notes of the AAS, 7, 220

\bibitem[{Gallardo(2024)}]{gallardo2024anthropogenic}
Gallardo, P.~A. 2024, Research Notes of the AAS, 8, 88

\bibitem[{Hajdukova {et~al.}(2020)Hajdukova, Sterken, Wiegert, \&
  Korno{\v{s}}}]{hajdukova2020challenge}
Hajdukova, M., Sterken, V., Wiegert, P., \& Korno{\v{s}}, L. 2020, Planetary
  and Space Science, 192, 105060

\bibitem[{Hunter(2007)}]{H07}
Hunter, J.~D. 2007, Computing In Science \& Engineering, 9, 90

\bibitem[{Jewitt \& Seligman(2023)}]{jewitt2023interstellar}
Jewitt, D. \& Seligman, D.~Z. 2023, Annual Review of Astronomy and
  Astrophysics, 61, 197

\bibitem[{Loeb(2022)}]{loeb2022possibility}
Loeb, A. 2022, Astrobiology, 22, 1392

\bibitem[{Loeb {et~al.}(2024{\natexlab{a}})Loeb, Adamson, Bergstrom, Cloete,
  Cohen, Conrad, Domine, Fu, Hoskinson, Hyung, {et~al.}}]{loeb2024recovery}
Loeb, A., Adamson, T., Bergstrom, S., {et~al.} 2024{\natexlab{a}}, arXiv
  preprint arXiv:2401.09882

\bibitem[{Loeb {et~al.}(2024{\natexlab{b}})Loeb, Adamson, Bergstrom, Cloete,
  Cohen, Conrad, Domine, Fu, Hoskinson, Hyung, {et~al.}}]{loeb2024recoveryb}
Loeb, A., Adamson, T., Bergstrom, S., {et~al.} 2024{\natexlab{b}}, Research
  Notes of the AAS, 8, 39

\bibitem[{Opik(1950)}]{opik1950interstellar}
Opik, E. 1950, Irish Astronomical Journal, vol. 1 (3), p. 80, 1, 80

\bibitem[{Pe{\~n}a-Asensio {et~al.}(2024{\natexlab{a}})Pe{\~n}a-Asensio,
  P.~Gr\`ebol-Tom\`as, Trigo-Rodr{\'\i}guez, Ram{\'\i}rez-Moreta, \&
  Kresken}]{epa2024iberian}
Pe{\~n}a-Asensio, E., P.~Gr\`ebol-Tom\`as, P., Trigo-Rodr{\'\i}guez, J.~M.,
  Ram{\'\i}rez-Moreta, P., \& Kresken, R. 2024{\natexlab{a}}, {\it submitted}
  [\eprint[arXiv]{astro-ph/2405.15024}]

\bibitem[{Pe{\~n}a-Asensio {et~al.}(2022)Pe{\~n}a-Asensio,
  Trigo-Rodr{\'\i}guez, \& Rimola}]{pena2022orbital}
Pe{\~n}a-Asensio, E., Trigo-Rodr{\'\i}guez, J.~M., \& Rimola, A. 2022, The
  Astronomical Journal, 164, 76

\bibitem[{Pe{\~n}a-Asensio {et~al.}(2024{\natexlab{b}})Pe{\~n}a-Asensio,
  Visuri, Trigo-Rodr{\'\i}guez, Socas-Navarro, Gritsevich, Siljama, \&
  Rimola}]{pena2024oort}
Pe{\~n}a-Asensio, E., Visuri, J., Trigo-Rodr{\'\i}guez, J.~M., {et~al.}
  2024{\natexlab{b}}, Icarus, 408, 115844

\bibitem[{P\'erez \& Granger(2007)}]{ipython07}
P\'erez, F. \& Granger, B.~E. 2007, Computing in Science and Engineering, 9, 21

\bibitem[{{Sheerin} \& {Loeb}(2021)}]{sheerin2021oumuamua}
{Sheerin}, T.~F. \& {Loeb}, A. 2021, Journal of the British Interplanetary
  Society, 74, 427

\bibitem[{Siegel(1982)}]{siegel}
Siegel, A.~F. 1982, Biometrika, 69, 242

\bibitem[{Siraj \& Loeb(2021)}]{siraj2021breakup}
Siraj, A. \& Loeb, A. 2021, Scientific Reports, 11, 3803

\bibitem[{Siraj \& Loeb(2022)}]{siraj2022meteor}
Siraj, A. \& Loeb, A. 2022, The Astrophysical Journal, 939, 53

\bibitem[{Siraj \& Loeb(2023)}]{siraj2023localizing}
Siraj, A. \& Loeb, A. 2023, Signals, 4, 644

\bibitem[{Socas-Navarro(2023)}]{socas2023candidate}
Socas-Navarro, H. 2023, The Astrophysical Journal, 945, 22

\bibitem[{Spearman(1904)}]{spearman1094proof}
Spearman, C. 1904, The American Journal of Psychology, 15, 72

\bibitem[{Tagliaferri {et~al.}(1994)Tagliaferri, Spalding, Jacobs, Worden, \&
  Erlich}]{tagliaferri1994detection}
Tagliaferri, E., Spalding, R., Jacobs, C., Worden, S.~P., \& Erlich, A. 1994,
  Hazards due to Comets and Asteroids, 24, 199

\bibitem[{Van Der~Walt {et~al.}(2011)Van Der~Walt, Colbert, \&
  Varoquaux}]{numpy11}
Van Der~Walt, S., Colbert, S.~C., \& Varoquaux, G. 2011, Computing in Science
  \& Engineering, 13, 22

\bibitem[{Virtanen {et~al.}(2020)Virtanen, Gommers, Oliphant, Haberland, Reddy,
  Cournapeau, Burovski, Peterson, Weckesser, Bright, {van der Walt}, Brett,
  Wilson, Millman, Mayorov, Nelson, Jones, Kern, Larson, Carey, Polat, Feng,
  Moore, {VanderPlas}, Laxalde, Perktold, Cimrman, Henriksen, Quintero, Harris,
  Archibald, Ribeiro, Pedregosa, {van Mulbregt}, Vijaykumar, Bardelli,
  Rothberg, Hilboll, Kloeckner, Scopatz, Lee, Rokem, Woods, Fulton, Masson,
  H{\"a}ggstr{\"o}m, Fitzgerald, Nicholson, Hagen, Pasechnik, Olivetti, Martin,
  Wieser, Silva, Lenders, Wilhelm, Young, Price, Ingold, Allen, Lee, Audren,
  Probst, Dietrich, Silterra, Webber, Slavin, Alfaro, {da Silva}, Vigna,
  Frederickson, Kelley, Mayer, Weichert, Bachant, {de Miranda Cardoso}, Reimer,
  Harrington, Rodriguez, Gusain, Tritz, Thoma, Caporta, Mohan, Pereira,
  Aldcroft, {Yadav}, Pingel, Robitaille, Spura, Jones, Cera, Leslie, Zito,
  Krauss, Upadhyay, Halchenko, V{\'a}zquez-Baeza, \&
  Contributors}]{2020SciPy-NMeth}
Virtanen, P., Gommers, R., Oliphant, T.~E., {et~al.} 2020, Nature Methods, 17,
  261

\end{thebibliography}

\end{document}